# The Historical Impact of Anthropogenic Climate Change on Global Agricultural Productivity

Ariel Ortiz-Bobea[1]*, Toby R. Ault[2], Carlos M. Carrillo[2], Robert G. Chambers[3], David B. Lobell[4].

**Affiliations:**

[1]Charles H. Dyson School of Applied Economics and Management, Cornell University, Ithaca, NY, USA.

[2]Department of Earth and Atmospheric Sciences, Cornell University, Ithaca, NY, USA.

[3]Department of Agricultural and Resource Economics, University of Maryland – College Park, College Park, MD, USA.

[4]Department of Earth System Science, Stanford University, Stanford, CA, USA.

*Correspondence to: ao332@cornell.edu.

**Abstract:**

Agricultural research has fostered productivity growth, but the historical influence of anthropogenic climate change on that growth has not been quantified. We develop a robust econometric model of weather effects on global agricultural total factor productivity (TFP) and combine this model with counterfactual climate scenarios to evaluate impacts of past climate trends on TFP. Our baseline model indicates that anthropogenic climate change has reduced global agricultural TFP by about 21% since 1961, a slowdown that is equivalent to losing the last 9 years of productivity growth. The effect is substantially more severe (a reduction of ~30-33%) in warmer regions such as Africa and Latin America and the Caribbean. We also find that global agriculture has grown more vulnerable to ongoing climate change.

**Main text**

Enhancing agricultural productivity is vital to lifting global living standards and advancing sustainable food production in the face of escalating challenges to agriculture and the environment (1-7). Investments in agricultural research have boosted agricultural productivity, but this growth in productivity has been distributed unequally across the world (8-10), and there are signs that it is slowing down (11-14). At the same time, human activities during the last century and a half caused global temperatures to rise by more than 1°C above their pre-industrial values (15). This increase affects the global weather patterns that are essential to agriculture (16-17). However, the impacts of this anthropogenic climate change (ACC) on the agricultural sector has not yet been quantified, as most research has focused on future impacts (18-19).

Research to date on the historical impact of ACC focuses overwhelmingly on yields of major cereal crops (20-21) or on total Gross Domestic Product (GDP) (22). However, recent studies in this area are of limited value for assessing overall agricultural productivity for the following reasons: (i) cereal crops represent only about 20% of agriculture's global net production value (Fig. S1), (ii) variations in measures such as yield, could deviate from changes in overall productivity if farmers also adjust inputs in response to weather (23-25), and (iii) growth and levels of total and agricultural GDP diverge considerably in most countries (26-28), and thus climate change impacts on total GDP could deviate considerably from agricultural impacts (29-30). There is much needed research on agricultural climate impacts beyond the effects on yields of the major staple crops (31).

We quantify the impact of ACC on global agricultural productivity since 1961. Instead of focusing on crop yield or agricultural output, we rely on a measure of agricultural Total Factor Productivity (TFP). TFP measures aggregate output per unit of *measured* aggregate input (32-35). TFP thus captures interactions between output and input adjustments that eluded earlier research. Here we rely on official TFP statistics for which agricultural output includes crops and livestock, while inputs encompass labor, land, physical capital and materials (12). However, these TFP statistics do not incorporate the effect of weather, as is generally the case.

Consider the production function $Y_{it} = e^{f(Z_{it})} A_{it} X_{it} U_{it}$, where $Y_{it}$ is aggregate agricultural output, $e^{f(Z_{it})}$ is the effect of weather $Z_{it}$, $A_{it}$ measures technological knowledge and $X_{it}$ and $U_{it}$ are the observed and unobserved aggregate inputs, respectively. The subscripts refer to individual countries ($i$) and year ($t$). The percentage change in TFP is approximated as

$$\Delta \ln TFP_{it} \equiv \Delta \ln(Y_{it}) - \Delta \ln(X_{it}) = \Delta \ln A_{it} + \Delta f(Z_{it}) + \Delta \ln U_{it}$$

where $\Delta$ denotes change. TFP growth reflects technological improvements embodied in $\Delta \ln A_{it}$, but also the unmeasured effects of random year-to-year weather changes $\Delta f(Z_{it})$ and unobserved input adjustments $\Delta \ln U_{it}$. While this aggregate representation may conceal fine-scale production processes that are important to practitioners in the field, it helps provide a much needed macro-economic picture about the global agricultural economy.

We ground this conceptual framework empirically by estimating an econometric model linking country-level TFP growth with weather change. Our model characterizes $f$ as a quadratic function of average temperature (T) and total precipitation (P) over the 5-month period centered around the greenest month of the year of each country or "green season" (See Methods):

$$\Delta \ln TFP_{it} = \alpha_i + \theta_t + \beta_1 \Delta T_{it} + \beta_2 \Delta T_{it}^2 + \beta_3 \Delta P_{it} + \beta_4 \Delta P_{it}^2 + \epsilon_{it}$$

Country-fixed effect $\alpha_i$ controls for average country TFP growth rates, and year-fixed effect $\theta_t$ for global shocks common to all nations. Conceptually, these parameters seek to control for technological change embodied in $\Delta \ln A_{it}$. Thus the $\beta$ coefficients are estimated via the within-country and within-year variation of TFP growth and year-to-year weather changes. The inclusion of squared terms $\Delta T_{it}^2$ and $\Delta P_{it}^2$ allows the effect of changes in weather to vary with baseline levels of $T$ or $P$. Unobserved changes in inputs that are not absorbed by $\alpha_i$ or $\theta_t$ and measurement errors in the TFP data are captured in the error term $\epsilon_{it}$. Note that measurement error in the TFP data (see Methods) that remain uncorrelated with year-to-year changes in weather do not introduce bias. We account for the uncertainty in the estimated parameters with a block bootstrap where we sample observations with replacement 500 times by year and region. We later consider more than 200 systematic variations of this model. For instance, we explore whether estimating separate response functions for various sub-regions of the world affects our results.

We summarize the key source data in Figure 1. The first panel, Fig. 1A, shows that average agricultural TFP has more than doubled since 1961 but that there is wide cross-country variation. Fig. 1B shows that there is a substantial range of variation in $\Delta \ln TFP_{it}$ across countries for any given year. We also observe that TFP has grown much more slowly in certain countries, particularly in Sub-Saharan Africa (Fig. 1C). In Fig. 1D we show how observed annual country-level average temperature (in circles) fall closer within the range of counterfactual weather trajectories with ACC (gold band) than without ACC (grey band). Fig. 1E shows analogous information for precipitation without any discernible pattern.

We find a robust relationship between agricultural TFP growth and weather changes (Fig. 2, Tab. S1). The temperature response function is roughly linear and downward sloping (Fig 2A) indicating that warmer temperatures over the green season are detrimental to TFP growth. We conduct two placebo checks that suggest this relationship is unlikely to arise by chance. These checks are based on the idea that incorrectly matching TFP growth and weather changes, and re-estimating, should yield results suggesting no or insignificant effects of weather on TFP. We first estimate models based on 10,000 "reshuffled" datasets that mismatch the *year* variable of the TFP growth and weather change data. The sample estimate (in Fig. 2A) falls outside the resulting distribution of spurious "reshuffled estimates" (Fig. 2B). In a second check we mismatch the *country* variable of the TFP growth and weather change data with similar results (Fig. 2C). The precipitation response function is non-linear and peaks at around 500 mm over the green season (Fig 2d). We demonstrate that this relationship is not likely spurious with the same placebo checks shown in Figs. 2E,F. We also find that the shape of the response functions is not driven by either hot or cold countries (Figs. S2, S3).

A critical question for climate change adaptation is whether agriculture is becoming more or less sensitive to climatic extremes. This would be reflected empirically as changes over time in the response functions shown in Fig. 2. We estimate a model based on the first (1962-1988) and second (1989-2015) halves of the sample and find that the temperature response function is noticeably steeper for the latter half (Figs. S4, S5). This indicates that higher temperatures have become more damaging. We formally confirm this by testing whether the temperature coefficient has changed between these two periods in a model with a linear specification for temperature ($P = 0.035$). We also find that the change in the temperature response function over time is not driven by isolated changes in outlying countries in the temperature distribution (Figs. S6-S9). This mirrors recent findings in US agriculture (36, 37).

We find no evidence that weather has a persistent effect on TFP growth. Our baseline specification only considers contemporaneous weather effects. But a weather shock could conceivably affect TFP growth in future years, for example if growth is faster following a year with bad weather. This would result in cumulative weather events affecting TFP growth. By introducing lags for weather in prior years, we cannot reject the hypothesis that the cumulative effect of changes in weather conditions up to 10 years in the past have no effect on TFP growth (Tab. S2). But rejecting this hypothesis may be challenging with aggregate data.

We subsequently link our econometric estimates with counterfactual weather trajectories from climate experiments with and without ACC to derive the cumulative impact of ACC for each country since 1961. We obtain the counterfactual weather trajectories from the Coupled Model Intercomparison Project Phase 6 (CMIP6). This approach combines both the statistical uncertainty from the econometric model regarding the climate-agriculture relationship and the climate uncertainty from the CMIP6 ensemble regarding the effect of human emissions on the climate system (see Methods).

The cumulative impact of ACC on global agricultural TFP growth over the 1961-2020 period is about $-20.8\%$ with a 90 percent confidence interval between $-36.2$ and $-11.0\%$ (Fig. 3A). In Fig. 3B, we represent this finding in levels by combining the counterfactual cumulative impacts of ACC on global TFP growth with the observed (1961-2015) and projected (2016-2020) global TFP level trajectory. This illustrates how much higher global TFP would have been without ACC. Specifically, we find that the global TFP level projected to be reached in 2020 in our world with ACC, would have been reached in 2011 in a world without ACC, with a 90 percent confidence interval between 2007 and 2015. That is, the impact of ACC represents a loss of the past 9 years of productivity growth.

This global result conceals sizeable regional and cross-country disparities. Figure 4 shows that the cumulative impact of ACC since 1961 is greater for warm regions like Africa ($-32.9\%$) and Latin American and the Caribbean ($-30.0\%$) than for cooler regions like North America ($-18.6\%$) and Europe and Central Asia ($-16.0\%$) (see Tab. S3). The large negative impacts for Africa seem particularly worrisome given the large portion of the population employed in agriculture. Overall, these findings are consistent with documented slowdowns in agricultural productivity (11-14), particularly in Sub-Saharan Africa (38,39) but also with studies analyzing economy-wide ACC impacts that exacerbate inequality between poor and rich countries (22, 40).

Our baseline global finding of the impact of ACC on global agricultural productivity is robust to 200 variations of the econometric model. Fig. 5 summarizes global estimates for the baseline and alternative models. The baseline model is shown in blue and corresponds to the estimate shown in Fig. 3A. Notice that using minimum (Tmin) and maximum (Tmax) temperature as alternative temperature variables or excluding precipitation does not substantially change our baseline estimate. In addition, using a cubic functional form to relax the symmetry of our baseline quadratic specification does not alter our baseline result. We also consider regressions with observations weighted by revenue and find those results to be systematically more pessimistic than our baseline model using equal weights. But aggregating weather data to the country level based on areas covered only by cropland or both cropland and pasture has little effect on our findings. We also consider models with weather variables aggregated over the entire calendar year. We find that those models fit less well and point to noticeably smaller damages. Finally, our baseline model imposes a single response function for the whole world. But allowing separate response functions for three equally-sized latitudinal groups of countries does not alter our findings. Overall, the 192 models that do not exclude observations point to an average mean impact of $-17.6\%$ with a standard deviation of $5.3\%$ (indicated with a horizontal red line and band in Fig. 5).

The exclusion of certain countries does not substantially affect our baseline estimate, but restricting the analysis to certain temporal subsamples does. A potential concern is that certain countries may overinfluence our estimates and thus drive our findings. But excluding certain large countries like China, the United States, India or Brazil, or the 10 percent coldest or hottest countries does not substantially alter our baseline finding (Fig. 5). However, basing our analysis on the latter part of the sample (1989-2015) points to even larger damages than our baseline estimate, on the order of $-30\%$ (Fig. 5). This reflects our finding that the response function is changing over time and suggests that global agriculture is growing increasingly sensitive to ongoing climate change.

Finally, we find that ignoring input responses by analyzing output rather than TFP, overstates the impact of ACC. The temperature response function is steeper for output than for TFP (Fig. S10), suggesting that farmers reduce aggregate input quantity in response to detrimental weather conditions. Ignoring confounding input adjustments, Fig. S11 indicates that ACC would have reduced output by about $32.3\%$ with a 90 percent confidence interval between $-61.3$ and $-16.8\%$. This is noticeably higher than the $20.8\%$ reduction based on TFP.

Our estimates should not be interpreted as the effect of a world without fossil fuels on global agricultural production. Agriculture has benefitted tremendously from agricultural research and carbon-intensive inputs that would not have been as available without fossil fuels. The counterfactual in our study only removes *the effect* that fossil fuels and other anthropogenic influences have on the climate system. For instance, our estimates do not remove the direct effect that rising $CO_2$ concentrations has on agricultural production or the presence of agricultural research or carbon-intensive inputs. In addition, the reader may be rightfully concerned about measurement error of our TFP metric. TFP estimates are notoriously difficult to construct, and require considerable background work sometimes using imperfect data sources. However, as long as the measurement error remains uncorrelated with year-to-year changes in TFP growth, which seems plausible, such errors do not bias our econometric estimates but simply render them more imprecise. Moreover, the TFP metric we rely upon likely mismeasures certain inputs, such as irrigation water use. Our own analysis shows that ignoring all measured inputs overstates ACC impacts, so the present analysis is a first step while more detailed international TFP statistics are produced. In conclusion, our study suggests that ACC is increasingly reducing agricultural output as we drift away from a climate system without human influences, cumulating into a detectable and sizeable impact as of 2020. Given recent productivity slowdowns (11-14), the long lags in agricultural research, and the rapid pace of ACC, our findings raise the question whether current levels of investments in agricultural research are sufficient to sustain 20[th] century rates of productivity growth in the 21[st] century.

**Methods**

*Data sources and data processing*

We obtain agricultural data from USDA ERS International Agricultural Total Factor Productivity (TFP) dataset (41). This dataset provides the most comprehensive set of international TFP estimates for the agricultural sector. The dataset provides country-level TFP indices (in levels) for 172 countries over the 1961-2015 period. The regression dataset has 9,255 observations with 172 countries and 54 years (1962-2015; note that one year is lost because of first differencing to compute growth rates). The dataset is balanced but for one nation, Palestine, for which we only have complete data from 1995. Note that some recent smaller nations were aggregated to their former larger countries to extend the time span of the dataset (e.g. Czechoslovakia, Yugoslavia, Ethiopia, Sudan, etc.).

TFP levels, annual growth rates and average growth rates are shown in Fig. 1A-C. The TFP growth rate for country $i$ and year $t$ is constructed by USDA as:

$$\Delta \ln TFP_{it} = \sum_j R_{itj} \ln \Delta Y_{itj} - \sum_k S_{itk} \ln \Delta X_{itk}$$

where $R_{itj}$ is the revenue share of the $j$th output $Y_{itj}$ and $S_{itk}$ is the cost share of the $k$th input $X_{itk}$. As indicated by USDA, TFP growth is the value-share-weighted difference between total output growth and total input growth. To avoid index number bias, the USDA adjusts weights $R_{itj}$ and $S_{itk}$ every decade. Outputs include crop and livestock commodities aggregated based on a common set of international prices derived by the Food and Agriculture Organization (FAO). Inputs include farm labor, agricultural land (quality adjusted), capital inputs (including farm machinery and livestock) and intermediate inputs (inorganic fertilizer and animal feed) which are mostly obtained from FAO. Thus changes in inventory in livestock herd size are taken into account. See ref (42) for more details. Note that although the data is constructed in terms of TFP growth rates, the data is transformed and reported in levels in the USDA website. Geographical delimitations in this study follow as close as possible FAO region definitions (Fig. S12). Note that the TFP metric we rely upon likely mismeasures certain inputs, such as irrigation water use. As discussed in ref (42), irrigation is accounted for in the land input through a quality adjustment of land equipped for irrigation. Thus changes in irrigation intensity remain unaccounted for. However, estimates of weather driven variability in global irrigation withdrawals remain modest at a global scale (~10%) (43).

We obtain the historical weather data from the Global Meteorological Forcing Dataset (GMFD) for land surface modeling developed by the Terrestrial Hydrology Research Group at Princeton University (44). The GMFD provides data on daily minimum and maximum temperature and total precipitation over 1948-2016 with a 0.25° spatial resolution (~28 km at the equator). Following standard practice in the literature, we aggregate these variables to the monthly level and then spatially aggregate the grids to the country level based on either cropland or cropland and pasture weights. We obtain these weights by resampling the 10km gridded land cover data in ref. 45 to the GMFD grid using a bilinear interpolation (Fig. S13). Although weighting weather data by production value may seem desirable, we find that 1- adopting alternative weighting schemes makes little practice difference, and 2- constructing a globally consistent set of grid-level production value weights would be prohibitive. We show the annual evolution of the country-level average temperature and percentage change in total precipitation for the 1961-2015 period in Fig. 1D,E.

We obtain the counterfactual monthly weather trajectories for average temperature and total precipitation for 1961-2020 from three sets of simulations in CMIP6. The "hist-nat" experiment (1961-2020) simulates the influence of natural forcing alone on the climate system. The "historical" experiment (1961-2014) simulates the influence of both human and natural forcings on the climate system. As stipulated by CMIP6, we complement this experiment with data for 2015-2020 from the SSP2-4.5 experiment. We rely (so far) on 7 GCMs (see Tab. S4).

To aggregate these modelled weather trajectories to the country-level using the same approach as above, we first downscale these data from their native GCM grid to the GMFD grid using the bias-corrected spatial disaggregation (BCSD) approach (46). BCSD corrects the bias of the modelled climate data and increases the spatial resolution with the ultimate goal of having a product of higher resolution that conserves the statistics of the global climate scenarios. The BCSD approach is performed in two steps. In the first step, we create a bias-corrected (BC) dataset by performing a quantile mapping to correct the bias (47, 48), which ref. 48 calls a "transformation". In the quantile mapping, we transform the GCM time series field to match the quantile distribution of the observed GMFD weather dataset, $Q_{GCM} \rightarrow Q_{OBS}$, using a common period for the transformation function (1961-2014). Because the transformation is done at every grid cell of the GCM, the observed GMFD dataset was aggregated to match the coarser GCM resolution. The approach was applied to both temperature and precipitation. In the second step, we increased the spatial resolution of the BC data by applying the spatial disaggregation (SD) approach. In the SD approach, we first removed the monthly observed climatology from the coarse resolution BC data, $\Delta F = F - CLIM_{COARSE}$. We then convert the anomaly to a high resolution by linear interpolation, $\Delta F \rightarrow \Delta F_{HIGH}$. Finally, we added the climatology with high resolution, $F_{HIGH} = \Delta F_{HIGH} + CLIM_{HIGH}$. Here, the anomaly calculation ($\Delta F$) is only valid for temperature. For precipitation the anomaly is computed using a ratio, $\Delta F = F/CLIM_{COARSE}$ with $F_{HIGH} = \Delta F_{HIGH} \times CLIM_{HIGH}$.

The baseline econometric model relies on weather variables aggregated over a 5-month period centered around the greenest month of year of each country based on Normalized Difference Vegetation Index (NDVI) climatology data (50). The NDVI data is the 3$^{rd}$ generation of NASA/GFSC GIMMS's NDVI dataset for 1981-2015. We first temporally aggregate the data to bi-weekly climatologies. We then smooth the climatology series within the year based on a 14-week moving window. We then identify the "greenest" month based on the month that includes the highest NDVI level of the year for each grid cell. The spatial distribution of the "greenest" month for each grid cell is shown in Fig. S14A. To obtain a country-level value, we first resample land cover weights to match that of the NDVI data. We then compute for each country the most frequent "greenest" month based on either cropland or cropland and pasture frequency weights. These country-level aggregations are shown in Fig. S14 B and C. For two small island nations (Fiji and Polynesia) there is no NDVI data. We therefore assign the greenest month to match that of the neighboring island nation of Vanuatu.

*Deriving the response function*

In order to help map our conceptual framework to the USDA TFP estimates, we consider the following relationship between aggregate output, aggregate input, weather, and technological knowledge $Y_{it} = e^{f(Z_{it})} A_{it} X_{it} U_{it}$, where $Y_{it}$ is aggregate agricultural output in country $i$ and year $t$, $e^{f(Z_{it})}$ is the effect of weather $Z_{it}$, $A_{it}$ measures current technological knowledge, and $X_{it}$ and $U_{it}$ are observed and unobserved aggregate inputs, respectively. By definition, TFP for country $i$ at time $t$ is $Y_{it}/X_{it}$ so that the percentage change in TFP for country $i$ at time $t$ is approximated as

$$\Delta \ln TFP_{it} \equiv \Delta \ln(Y_{it}) - \Delta \ln(X_{it}) = \Delta \ln A_{it} + \Delta f(Z_{it}) + \Delta \ln U_{it}$$

Empirically, our econometric models seek to control for $\Delta \ln A_{it}$ through country and year fixed effects ($\alpha_i$ and $\theta_t$ respectively) and model $\Delta f(Z_{it})$ in various ways. Because the model is specified in growth terms, the inclusion of a country-specific dummy variable $\alpha_i$ is analogous to controlling for a linear country-specific time trend in $\ln TFP$. Unobserved inputs that are not absorbed by the fixed effects are captured in the error term $\epsilon_{it}$. This error term also captures measurement errors in the TFP metric, which includes changes in irrigation water use not fully captured by changes in irrigated area. Perhaps with the exception of water withdrawals, measurement errors in the TFP data are unlikely to be correlated with year-to-year weather changes, which does not bias our results. Our baseline model regresses $\Delta \ln TFP_{it}$ on first differences of green-season average temperature and precipitation:

$$\Delta \ln TFP_{it} = \beta_1 \Delta T_{it} + \beta_2 \Delta T_{it}^2 + \beta_3 \Delta P_{it} + \beta_4 \Delta P_{it}^2 + \alpha_i + \theta_t + \epsilon_{it}$$

To capture the statistical uncertainty of the regression model we conduct a block bootstrap estimation where we sample observations by year-region with replacement. While there is serial dependence in TFP levels, previous work has shown there is no serial dependence in growth $\Delta \ln TFP_{it}$ (e.g. 14, 36). We thus focused on accounting for contemporaneous regional dependence. Regions correspond to the seven FAO regions shown in Fig. S12. We show the response function with a 90% bootstrapped confidence band for temperature and precipitation in Figs. 2A, D. We show regression coefficients for the baseline model in Tab. S1. Weather parameters $\beta$ are subsequently used in a simulation to derive the effect of ACC on global agricultural TFP. Importantly, note that measurement error in $\Delta \ln TFP_{it}$ would need to be correlated with $\Delta T_{it}$ and/or $\Delta P_{it}$ to induce any form of bias in the estimation of $\hat{\beta}$. In addition, classical measurement error in $\Delta T_{it}$ and/or $\Delta P_{it}$ would induce attenuation bias, reducing the magnitude of our findings.

The placebo checks shown in Figs. 2B, C, E, F evaluate whether the estimated relationship is spurious. The idea is to evaluate the chances that the result is spurious by contrasting the estimated coefficients in our sample with a distribution of coefficients from "reshuffled" datasets where we should, on average, expect no effect of weather. We perform 10,000 regressions based on datasets that are mismatched by year and by country. In all cases the estimated coefficients fall clearly outside the distribution of "reshuffled" estimates, in support of our baseline model.

*Robustness checks*

A crucial concern in applied econometric analysis is that baseline models proposed by researchers may not be robust to even small variations in model specification or the underlying data. We conduct a systematic exploration spanning 200 variations of the econometric model to assess the robustness of our baseline finding (Fig. 5).

We consider all possible combinations of models along the following dimensions: 1- relies on either Tmax, Tmin or Tmean, 2- includes or excludes precipitation, 3- adopts a quadratic or cubic functional form for all weather variables, 4- relies on equal or revenue regression weights, 5- relies on weather data aggregated over cropland or cropland and pasture, 6- relies on a the calendar year or the "green season" (5 months centered around the greenest month) for aggregating weather conditions, and 7- relies on a single global response function or on separate response functions for 3 equally-sized latitudinal regions. This corresponds to 192 variations of the model.

We find most these variations have relatively little bearing on the baseline result presented in the paper in Fig. 3. The adoption of revenue regression weights and a "green season" for weather aggregation represent the two most consequential modeling choices, pointing to larger damages.

We also consider 8 data restrictions including country exclusions (China, USA, India, Brazil, coldest 10%, hottest 10%) and temporal subsets (1962-1988 and 1989-2015) which are discussed in the main paper.

*Measurement error*

Measurement error affects all econometric analyses and its consequences are well understood (51). Here we characterize how it might affect our study. As stated earlier, our dependent variable $\Delta \ln TFP_{it}$ is possibly mismeasured. Measurement error in $\Delta \ln TFP_{it}$ that is uncorrelated with weather (classical measurement error) does not introduce bias in the estimation but renders our less precise. Only measurement error in $\Delta \ln TFP_{it}$ that is correlated with weather fluctuations (non-classical measurement error) introduces bias. It seems unlikely that weather may affect the collection and reporting of output and input quantities. Thus systematic differences across country in data reporting would not introduce bias.

However, omitted or mismeasured variable input adjustments in response to weather fluctuations could be problematic. The USDA ERS International Agricultural Total Factor Productivity dataset (39) we rely upon accounts for irrigation via the amount of land equipped for irrigation (40), not through water withdrawals directly. If farmers increase groundwater irrigation intensity in response to unfavorable conditions, but such short term increases in irrigation are overlooked, then unfavorable weather conditions would appear less damaging. Note that surface irrigation (e.g. flood irrigation) may be procyclical with precipitation, so the overall relationship between irrigation intensity and weather fluctuations seems indeterminate. Our analysis focusing on output (rather than TFP) suggests that ignoring input responses affects the estimation of the sensitivity to weather conditions (Fig. S10). But the overall role of irrigation may be limited because weather-driven variability in global irrigation withdrawals remains modest at a global scale (~10%) (43). Moreover, our main global finding remains stable when we estimated the response function separately for various regions of the world (Fig. 5).

Another potential concern is measurement error in our independent weather variables. Classical measurement error in independent variables, here $\Delta T_{it}$ and $\Delta P_{it}$, cause attenuation bias (bias toward zero). This means that such errors would tend to "dilute" our findings toward no results. On the other hand, measurement error that is correlated with weather fluctuations could bias our findings in either direction. However, comparisons of regression results across alternative datasets in other studies generally show minimal discrepancies.

*Impact of anthropogenic climate change*

We compute the impact of anthropogenic climate change on each country's agricultural TFP growth by subtracting the cumulative impact of a weather trajectory with ACC for a given GCM from the cumulative impact of a weather trajectory without ACC for the same GCM. For a country $i$, the cumulative impact from 1962 to year $t_0$ for a weather trajectory (with or without ACC) is computed as:

$$I_{it_0} = \sum_{t=1962}^{t_0} \Delta \ln \widehat{TFP}_{it} = \widehat{\beta_1} \sum_t \Delta T_{it} + \widehat{\beta_2} \sum_t \Delta T_{it}^2 + \widehat{\beta_3} \sum_t \Delta P_{it} + \widehat{\beta_4} \sum_t \Delta P_{it}^2$$

where the changes in weather variables (e.g. $\Delta T_{it}$) are the differences between the sequence of seasonal weather conditions relative to the 23-year climatology centered around 1961 (1950-1972) for the scenario without ACC for that particular GCM. Thus, the cumulative impact of ACC is $I_{it_0}^{\text{with ACC}} - I_{it_0}^{\text{without ACC}}$ in year $t_0$. We can compute this cumulative impact for all years between 1962 and 2020 for a particular set of values of the $\hat{\beta}$ coefficients and a GCM.

To reflect the joint statistical uncertainty from the econometric model and climate uncertainty arising from various GCMs in CMIP6, we compute cumulative impacts of ACC for 2,000 random pairs of bootstrapped coefficients $\hat{\beta}$ and GCMs. Figure 3A shows the 2,000 trajectories of the cumulative impact of ACC for all years in 1962-2020, as well as the distribution of those impacts on 2020, on the right.

In Fig. 3B we illustrate the impact of ACC by contrasting counterfactual TFP level trajectories with the observed TFP level trajectory. Specifically, we obtain a counterfactual TFP level trajectory $L_{it_0}$ for country $i$ at year $t_0$ by taking the exponential of the observed TFP level trajectory (mostly $> 0$) minus each one of the 2,000 counterfactual cumulative impacts of ACC (mostly $< 0$):

$$L_{it_0} = \exp\left( \underbrace{\sum_{t=1962}^{t_0} \Delta \ln TFP_{it}^{\text{observed}}}_{\text{Observed TFP level trajectory}} - \underbrace{\left( I_{it_0}^{\text{with ACC}} - I_{it_0}^{\text{without ACC}} \right)}_{\substack{\text{Counterfactual cumulative} \\ \text{impact of ACC}}} \right)$$

These counterfactual TFP level trajectories are shown in grey with a blue line and band showing their mean and 90% confidence intervals, respectively. The red solid line shows the observed TFP level trajectory, $\exp\left(\sum_{t=1962}^{t_0} \Delta \ln TFP_{it}^{\text{observed}}\right)$, for the 1962-2015 period. Because the TFP dataset extends only to 2015, we project the TFP trajectory for 2016-2020 (shown in a dashed red line) based on the average growth rate over the previous 10 years (2006-2015) for each country. Regional and global cumulative impacts of ACC are obtained by aggregating country-level cumulative impacts based on the fixed revenue weights for each country, and then converting these to levels after regional aggregation. TFP level trajectories are normalized to 100 in 1962.

**References and Notes:**

1.  Johnston, B.F., & Mellor, J.W. The role of agriculture in economic development. *The American Economic Review* **51**, 566-593 (1961).

2.  Timmer, C.P. The agricultural transformation. *Handbook of development economics* **1**, 275-331 (1988).

3.  Barrett, C.B., Carter, M.R. & Timmer, C.P. A century-long perspective on agricultural development. *American Journal of Agricultural Economics* **92**, 447-468 (2010).

4.  de Janvry, A., & Sadoulet, E. Agricultural growth and poverty reduction: Additional evidence. *The World Bank Research Observer* **25**, 1-20 (2010).

5.  Christiaensen, L., Demery, L., & Kuhl, J. The (evolving) role of agriculture in poverty reduction—An empirical perspective. *Journal of Development Economics* **96**, 239-254 (2011).

6.  World Bank. World Development Report 2008: Agriculture for Development. Washington, DC. (2007). https://openknowledge.worldbank.org/handle/10986/5990.

7.  Curtis, P.G., Slay, C.M., Harris, N.L., Tyukavina, A., & Hansen, M.C. Classifying drivers of global forest loss. *Science* **361**, 1108-1111 (2018).

8.  Steensland, A. 2019 Global Agricultural Productivity Report: Productivity Growth for Sustainable Diets and More (2019). http://hdl.handle.net/10919/96429

9.  International Food Policy Research Institute (IFPRI). 2020 Global Food Policy Report: Building Inclusive Food Systems. Washington, DC: International Food Policy Research Institute (IFPRI). (2020) https://doi.org/10.2499/9780896293670

10. Fuglie, K. R&D capital, R&D spillovers, and productivity growth in world agriculture. *Applied Economic Perspectives and Policy*, **40**, 421-444 (2018).

11. Alston, J.M., Beddow, J.M., & Pardey, P.G. Agricultural research, productivity, and food prices in the long run. *Science* **325**, 1209-1210 (2009).

12. Fuglie, K. O., Wang, S. L., & Ball, V. E. (Eds.). Productivity growth in agriculture: an international perspective. CABI (2012).

13. Ball, V.E., Sheng, Y., Mesonada, C., & Nehring, R. Comparing Agricultural TFP between 17 OECD Countries, 1973-2011, *Unpublished Manuscript* (2018).

14. Chambers, R.G., Pieralli, S., & Sheng, Y. The Millennium Droughts and Australian Agricultural Productivity Performance: A Nonparametric Analysis. *American Journal of Agricultural Economics* (2019).

15. IPCC (2014). Climate change 2014: Synthesis report. In R. K. Pachauri & L. A. Meyer (Eds.), (151 pp.) Geneva, Switzerland: IPCC.

16. Lesk, C., Rowhani, P. & Ramankutty, N. Influence of extreme weather disasters on global crop production. *Nature* **529**, 84-87 (2016).

17. Ray, D.K., Gerber, J.S., MacDonald, G.K., & West, P.C. Climate variation explains a third of global crop yield variability. *Nature Communications* **6**, 1-9 (2015).

18. Zhao, C., Liu, B., Piao, S., Wang, X., Lobell, D. B., Huang, Y., ... & Durand, J. L. Temperature increase reduces global yields of major crops in four independent estimates. *Proceedings of the National Academy of Sciences*, **114**, 9326-9331 (2017).

19. Liu, B., Asseng, S., Müller, C., Ewert, F., Elliott, J., Lobell, D. B., ... & Rosenzweig, C. Similar estimates of temperature impacts on global wheat yield by three independent methods. *Nature Climate Change*, **6**, 1130-1136 (2016).

20. Lobell, D.B., & Field, C.B. Global scale climate–crop yield relationships and the impacts of recent warming. *Environmental Research Letters* **2**, 014002 (2007).

21. Lobell, D.B., Schlenker, W., & Costa-Roberts, J. Climate trends and global crop production since 1980. *Science* **333**, 616-620 (2011).

22. Diffenbaugh, N. S., & Burke, M. Global warming has increased global economic inequality. *Proceedings of the National Academy of Sciences*, ***116***, 9808-9813 (2019).

23. Kaufmann, R.K., & Snell, S.E. A biophysical model of corn yield: integrating climatic and social determinants. *American Journal of Agricultural Economics* **79**, 178-190 (1997).

24. Jagnani, M., Barrett, C. B., Liu, Y., & You, L. Within-Season Producer Response to Warmer Temperatures: Defensive Investments by Kenyan Farmers. *The Economic Journal*. Forthcoming (2020).

25. Aragón, F. M., Oteiza, F., & Rud, J. P. Climate Change and Agriculture: Subsistence Farmers' Response to Extreme Heat. *American Economic Journal: Economic Policy*. Forthcoming (2020).

26. Schultz, T. W. Transforming traditional agriculture. *Yale University Press*. New Haven (1964).

27. Gollin, D., Lagakos, D., & Waugh, M.E. The agricultural productivity gap. *The Quarterly Journal of Economics* **129**, 939-993 (2014).

28. Adamopoulos, T., & Restuccia, D. The size distribution of farms and international productivity differences. *American Economic Review* **104**, 1667-97 (2014).

29. Dell, M., Jones, B.F. & Olken, B.A. Temperature shocks and economic growth: Evidence from the last half century. *American Economic Journal: Macroeconomics* **4**, 66-95 (2012).

30. Burke, M., Hsiang, S.M. & Miguel, E. Global non-linear effect of temperature on economic production. *Nature* **527**, 235-239 (2015).
13

31. Hertel, Thomas W., and Cicero Z. de Lima. "Climate impacts on agriculture: Searching for keys under the streetlight." *Food Policy* (2020): 101954.

32. Comin, D. Total factor productivity. *Economic growth*, 260-263. Palgrave Macmillan, London (2010).

33. Hulten, C.R. Total factor productivity: a short biography. In *New developments in productivity analysis*, pp. 1-54. University of Chicago Press (2001).

34. Van Beveren, I. Total factor productivity estimation: A practical review. *Journal of economic surveys* **26**, 98-128 (2012).

35. Coomes, O. T., Barham, B. L., MacDonald, G. K., Ramankutty, N., & Chavas, J. P. (2019). Leveraging total factor productivity growth for sustainable and resilient farming. *Nature Sustainability*, *2*(1), 22-28.

36. Liang, X-Z, Wu, Y., Chambers, R.G., Schmoldt, D.L., Gao, W., Liu, C., Liu, Y-A, Sun, C., & Kennedy, J.A. Determining climate effects on US total agricultural productivity. *Proceedings of the National Academy of Sciences* **114**, E2285-E2292 (2017).

37. Ortiz-Bobea, A., Knippenberg, E., & Chambers, R.G. Growing climatic sensitivity of US agriculture linked to technological change and regional specialization. *Science Advances* **4**, eaat4343 (2018).

38. Frisvold, G., & Ingram, K. (1995). Sources of agricultural productivity growth and stagnation in sub-Saharan Africa. *Agricultural Economics*, *13*(1), 51-61.

39. Fuglie, K., & Rada, N. (2013). Resources, policies, and agricultural productivity in sub-Saharan Africa. *USDA-ERS Economic Research Report*, (145).

40. Letta, M., & Tol, R.S.J. Weather, climate and total factor productivity. *Environmental and Resource Economics* **73**, 283-305 (2019).

41. United States Department of Agriculture Economic Research Service. 2019. International Agricultural Productivity. Online database. Washington, DC (accessed June 19, 2020). https://www.ers.usda.gov/data-products/international-agricultural-productivity/

42. Fuglie, K. Accounting for growth in global agriculture. *Bio-based and Applied Economics* **4**, 221-254 (2015).

43. Wisser, D., Frolking, S., Douglas, E. M., Fekete, B. M., Vörösmarty, C. J., & Schumann, A. H. Global irrigation water demand: Variability and uncertainties arising from agricultural and climate data sets. *Geophysical Research Letters*, *35*(24) (2008).

44. Sheffield, J., G. Goteti, and E. F. Wood. Development of a 50-yr high-resolution global dataset of meteorological forcings for land surface modeling, *J. Climate*, 19 (13), 3088-3111 (2006). https://hydrology.princeton.edu/data.pgf.php

45. Ramankutty, N., Evan, A.T. , Monfreda, C., & Foley, J.A. Farming the planet: 1. Geographic distribution of global agricultural lands in the year 2000. *Global Biogeochemical Cycles* **22**, GB1003 doi:10.1029/2007GB002952 (2008).

46. Wood, A.W., Leung, L.R., Sridhar, V., Lettenmaier, D.P. Hydrologic implications of dynamical and statistical approaches to downscaling climate model outputs. *Climatic Change* **62**, 189-214 (2004).

47. Li, H., Sheffield, J., Wood, F.E. Bias correction of monthly precipitation and temperature fields from Intergovernmental Panel on Climate Change AR4 models using equidistant quantile matching. *Journal of Geophysical Research* **115**, D10101, doi:10.1029/2009JD012882 (2010).

48. Maurer, E.P., Ficklin, D.L, Wang, W. The impact of spatial scale in bias correction of climate model output for hydrologic impacts studies. *Hydrology and Earth System Science* **20**, 685-696 (2016).

49. Panofsky, H.A., Brier, G.W. Some applications of statistics to meteorology. The Pennsylvania State University, University Park, 224 (1963).

50. National Center for Atmospheric Research Staff (Eds). Last modified 14 Mar 2018. "The Climate Data Guide: NDVI: Normalized Difference Vegetation Index-3rd generation: NASA/GFSC GIMMS." Retrieved from https://climatedataguide.ucar.edu/climate-data/ndvi-normalized-difference-vegetation-index-3rd-generation-nasagfsc-gimms.

51. Hausman, J. Mismeasured variables in econometric analysis: problems from the right and problems from the left. *Journal of Economic perspectives*, *15*(4), 57-67 (2001).

**Acknowledgments:** We thank Chris B. Barrett and participants at the AERE and EAAE summer meetings, Giannini Foundation's Big Ag Data Conference, and at seminars at Cornell University, Arizona State University, University of Arizona, North Carolina State University and Duke University for useful comments. **Funding:** AOB was partially supported by the USDA National Institute of Food and Agriculture, Hatch/Multi State project 1011555. TRA and CMC were partially supported by NSF grants 1602564 and 1751535, as well as the Atkinson Center for a Sustainable Future, the Cornell Initiative for Digital Agriculture, and the Braudy Foundation. **Author contributions:** AOB conceived the study, conducted and led research and the writing of the manuscript. CMC obtained and downscaled modelled climate data. TRA, RGC, DBL provided detailed guidance and advice throughout the project. All authors contributed to writing the manuscript. **Competing interests:** Authors declare no competing interests. **Data and materials availability:** All data and code necessary to fully reproduce results in this study are deposited in a permanent online repository at the Cornell Institute for Social and Economic Research (CISER): https://doi.org/10.6077/pfsd-0v93.

**Supplementary Information:**

Figures S1-S14

Tables S1-S4

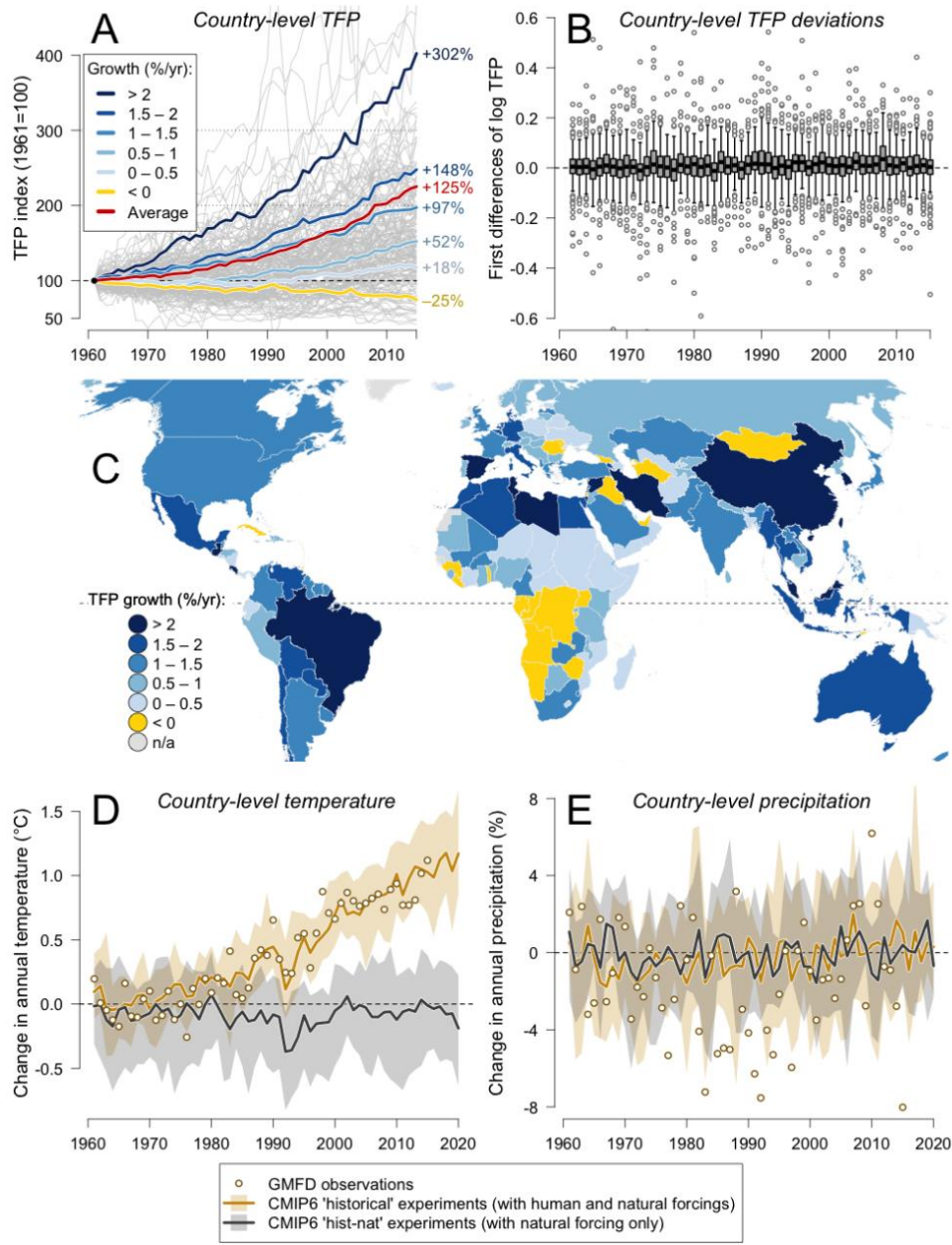

**Fig. 1: Recent trends in agricultural productivity and climate.** (**A**) Country-level growth in agricultural Total Factor Productivity (TFP) over 1961-2015. Grey lines indicate observed TFP level trajectories for all countries in the sample. Colored lines correspond to the average TFP level trajectories for countries with varying average TFP growth rates. (**B**) Distribution of first differences in the log of country-level TFP. The boxes represent the first 3 quartiles (Q1, Q2, Q3). Whiskers extend to 1.5 times the interquartile range (IQR=Q3-Q1). Observations falling beyond 1.5 IQR are represented with small circles. (**C**) Map representing the annual average growth rate in agricultural TFP over 1961-2015. (**D** and **E**) Evolution of global average annual temperature and annual precipitation of the Global Meteorological Forcing Dataset (GMFD) observations (circles). The golden band extends to the range of modelled variables from CMIP6 (7 GCMs). Simple averages of country-level variables are shown, thus small countries are over-represented.

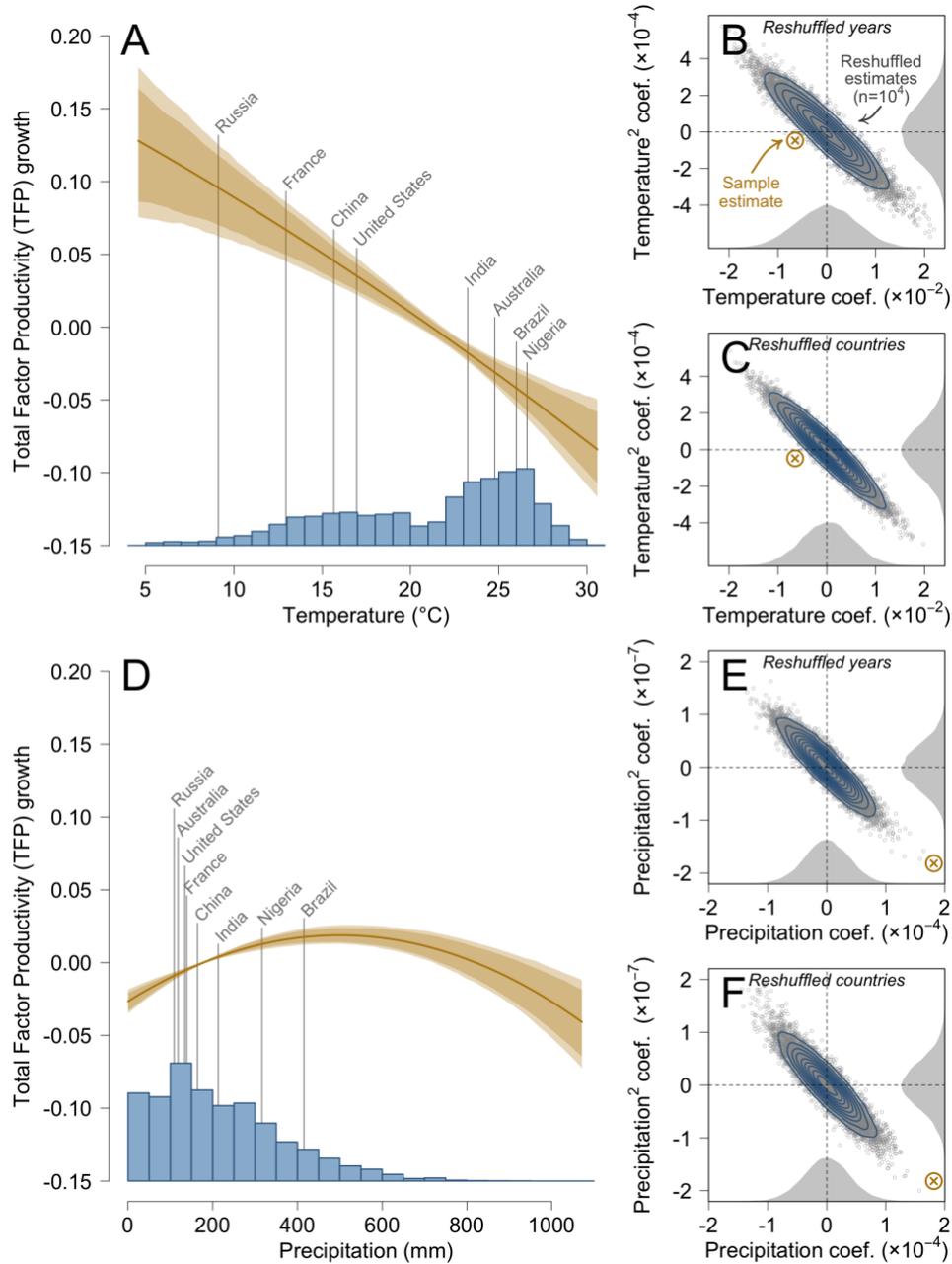

**Fig. 2: The response of agricultural productivity to weather.** (**A**) Response function of TFP growth to changes in green-season average T. Response functions are centered vertically so that the exposure-weighted marginal effect is zero. The colored bands represent 90 and 95% confidence bands based on 500 year-by-region block bootstraps. The blue bars represent the country-level distribution of green-season average T over the sample period 1962-2015. The average green-season T is indicated for a select number of large countries. (**B**) Panel shows the result of a placebo check whereby TFP and weather data are randomly mismatched or reshuffled by years. The distribution represents the linear and quadratic T coefficients based on 10,000 reshuffled datasets. (**C**) Same as previous panel but based on datasets reshuffled by country. (**D**) Response function of changes in country-level TFP to changes in green-season total P. (**E**) Same as panel **B** but for P coefficients. (**F**) Same as panel **C** but for P coefficients.

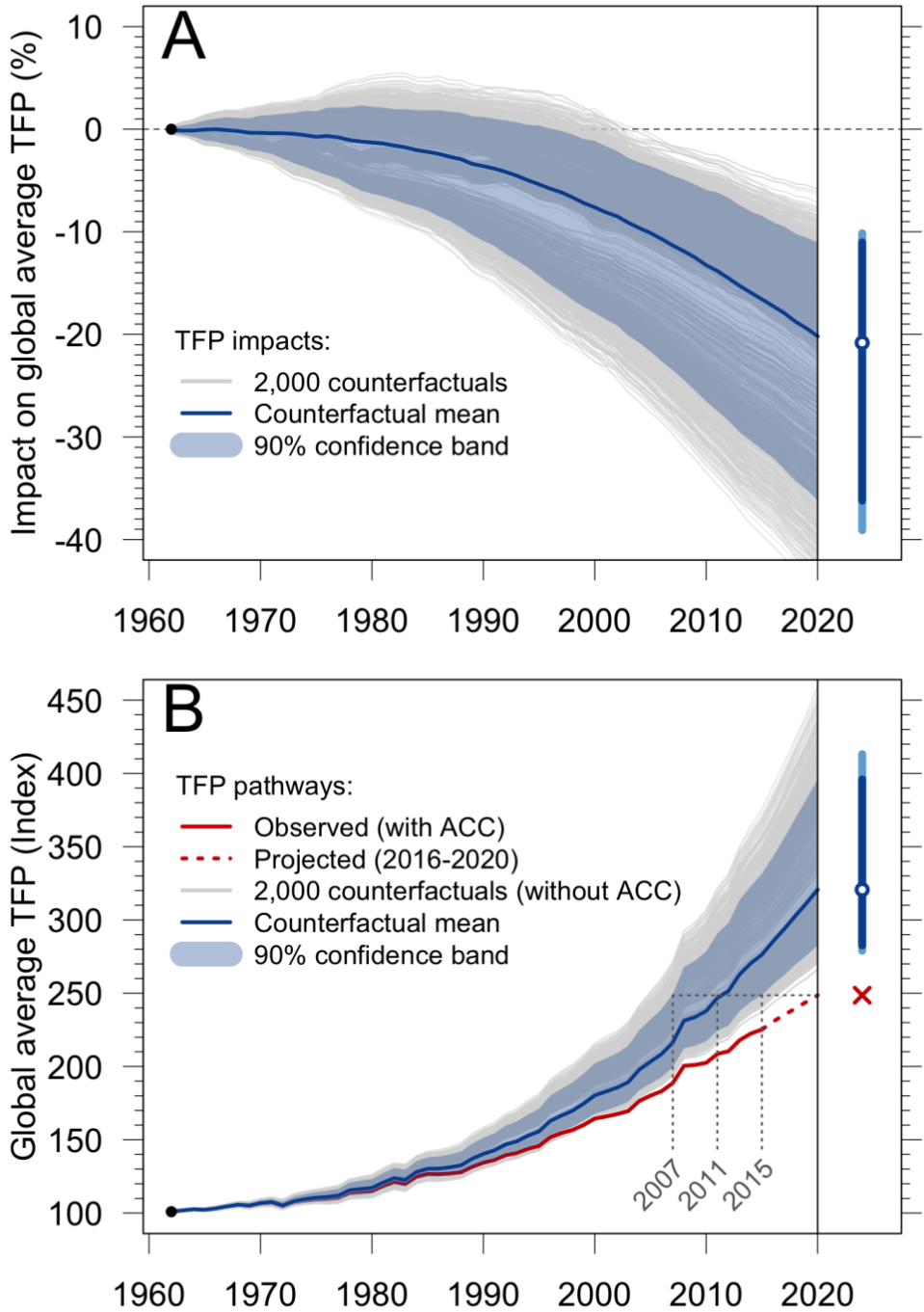

**Fig. 3: Global impact of anthropogenic climate change on productivity.** (**A**) 2,000 counterfactual paths ways combining statistical uncertainty and climate uncertainty. Blue line shows ensemble mean and the blue cone represents a 90 percent confidence band. The error bars on the right indicate 90 and 95% confidence intervals for the impact on 2020. (**B**) Same results presented in levels relative to the observed (1962-2015) and projected (2016-2020) level of TFP (in red).

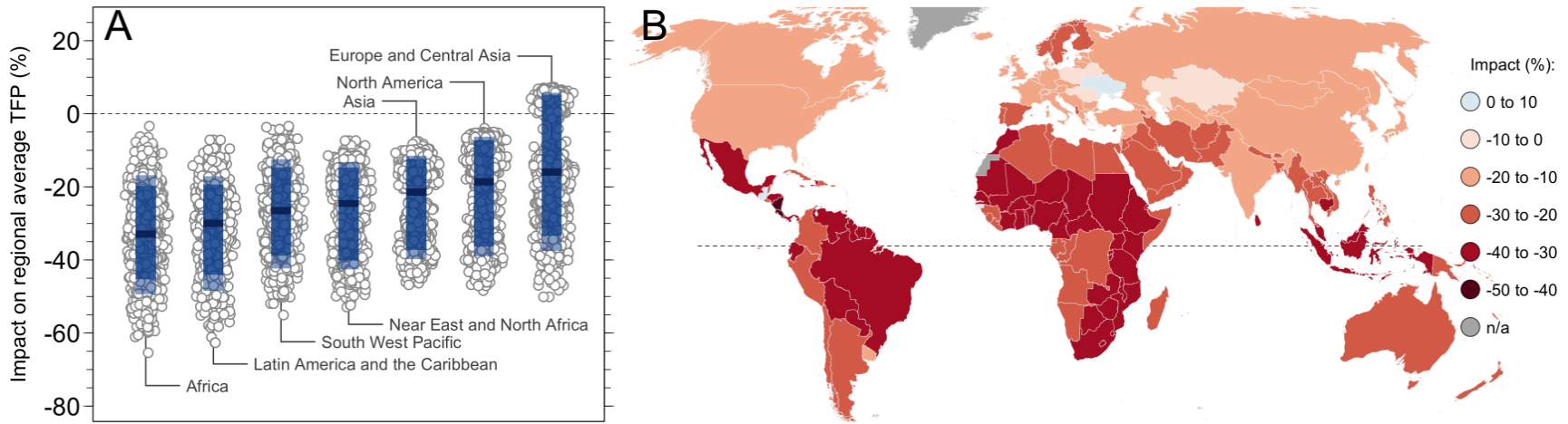

**Fig. 4: Regional and country-level impacts of anthropogenic climate change.** (**A**) Impact estimates for the baseline mode for each region. The white circles represent 2,000 estimates for each region. The blue bars represent 90 and 95% confidence bands and the solid line indicates the ensemble mean. (**B**) The color corresponds to the ensemble mean impact for each country in the sample.

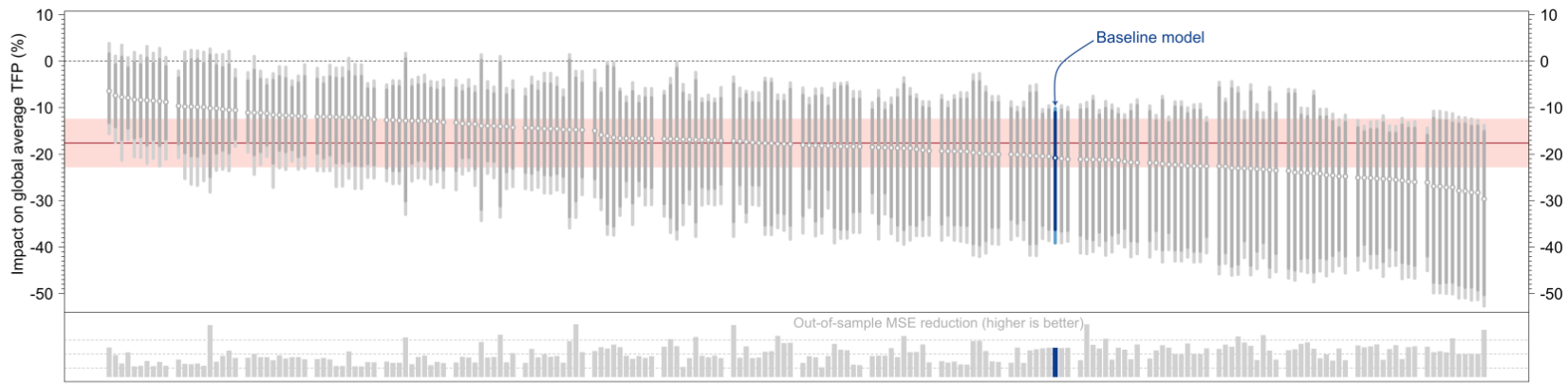

**Fig. 5: Global impact of anthropogenic climate change under multiple econometric models.** Vertical lines represent the 90 and 95% confidence intervals (in light and dark color, respectively) around the ensemble mean estimate for a particular model. ACC impacts for the baseline model, also shown in Fig. 3A, is highlighted in blue whereas alternative models are shown in grey. The red horizontal line and band represent the average mean impact of the 192 models out of the 200 that do not exclude observations, plus and minus a standard deviation ($-17.6 \pm 5.3\%$). The vertical bars directly below the impact estimates represent the reduction in out-of-sample MSE of a 10-fold cross-validation (whereby years of data are sampled together) relative to a model that excludes weather variables. Thus, higher bars indicate better model fit. The dotted table below provides information about the characteristics of each econometric model.